\documentclass[12pt]{article}

\usepackage{tikz}
\usepackage{amsmath}

\usepackage{makeidx,amssymb,amsmath,amsthm,graphicx}

\pdfoutput=1

\begin{document}


\title{Single crystal growth and characterization of GdRh$_2$Si$_2$}


\author{K.~Kliemt, C.~Krellner} 
 \date{Kristall- und Materiallabor, Physikalisches Institut, 
 Goethe-Universit\"at Frankfurt, Max-von-Laue Stasse 1, \\
 60438 Frankfurt am Main, Germany\\
\vspace{0.5cm}
kliemt@physik.uni-frankfurt.de}
\maketitle
\begin{abstract}
High-temperature indium flux growth was applied to prepare 
single crystals of GdRh$_2$Si$_2$ by a modified Bridgman method leading to mm-sized single crystals with a platelet habitus. 
Specific heat and susceptibility data of GdRh$_2$Si$_2$ exhibit a pronounced anomaly at $T_N = 107\rm \,K$, where the AFM ordering sets in.
Magnetic measurements on the single crystals were performed down to $T = 2$\,K in external fields from B = 0 - 9\,T applied along the 
$[100]$-, $[110]$- and $[001]$-direction of the tetragonal lattice. 
The effective magnetic moment determined from a Curie-Weiss fit 
agrees well with values from literature, and is larger than the theoretically predicted value.  Electrical transport data recorded for current flow parallel and perpendicular to the $[001]$-direction show a large anisotropy below $T_N$. The residual resistivity ratio $\rm RRR=\rho_{300K}/\rho_{0}\sim 23$ demonstrates that we succeeded in preparing high-quality crystals using high-temperature indium flux-growth.
\end{abstract}



\section{Introduction}
\label{}
Among the ternary silicides of the type RT$_2$Si$_2$ (R = rare earth, T = transition metal) which crystallize in the body-centered tetragonal ThCr$_2$Si$_2$ structure, GdRh$_2$Si$_2$ has attracted much attention in the last decades as it belongs to the compounds with rare earth elements with exceptional magnetic properties, e.g., CeRh$_2$Si$_2$ \cite{quezel1984}, 
YbRh$_2$Si$_2$ \cite{trovarelli2000} and EuRh$_2$Si$_2$ \cite{seiro2014, chikina2014}.
Studies of Gd-compounds are of special interest since the $4f$ shell of Gd 
is half filled and therefore its ground state with $S=7/2$ and $L=0$ 
is insensitive to crystal-electric-field (CEF) effects. In the past, polycrystalline GdRh$_{2}$Si$_{2}$  was subject 
to several investigations.
Magnetization and M\"ossbauer studies were performed 
by Felner and Nowik \cite{felner1983, felner1984}
examining the series RRh$_2$Si$_2$ and by Czjzek \textit{et al.} \cite{czjzek1989} with focus on the properties of the transition metal T
in the compounds GdT$_2$Si$_2$.
From M\"ossbauer spectra of $^{155}$Gd 
in GdRh$_2$Si$_2$ it was deduced that the rare earth local moments order antiferromagnetically 
with the ordering in the basal plane perpendicular to the fourfold symmetry axis of the tetragonal lattice
\cite{felner1983, felner1984}. It is known from neutron diffraction experiments, 
that the antiferromagnetic properties in the series RRh$_2$Si$_2$ arise due to a stacking of ferromagnetic layers \cite{slaski1983}.
Pressure studies by Szytu{\l}a \textit{et al.} \cite{szytula1986} revealed that the N\'eel temperature of GdRh$_{2}$Si$_{2}$ decreases with increasing applied pressure. 
An ESR-study was performed by Kwapulinska \textit{et al.} \cite{kwapulinska1988} and they found that the $g$-factor is temperature independent from $T_N$ to 300\,K with $g=1.995\pm0.01$.
Recently, the magnetic properties of GdRh$_2$Si$_2$ were investigated 
by hyperfine interactions and magnetization measurements \cite{Cabrera2012}.   
\\
All measurements up to now were carried out on polycrystalline material since single crystals were 
not available. Here, we report on the successful crystal growth and present a detailed study of the magnetic 
and electrical transport properties of GdRh$_2$Si$_2$ single crystals.
The crystals were grown using a high-temperature indium-flux technique \cite{Canfield, Canfield2001}.
The crystal growth set up was chosen similarly to that applied for the 
single crystal growth of YbRh$_2$Si$_2$ \cite{krellner2012}.

\section{Experimental details}

\begin{figure}
\centering
\includegraphics[width=0.7\textwidth]{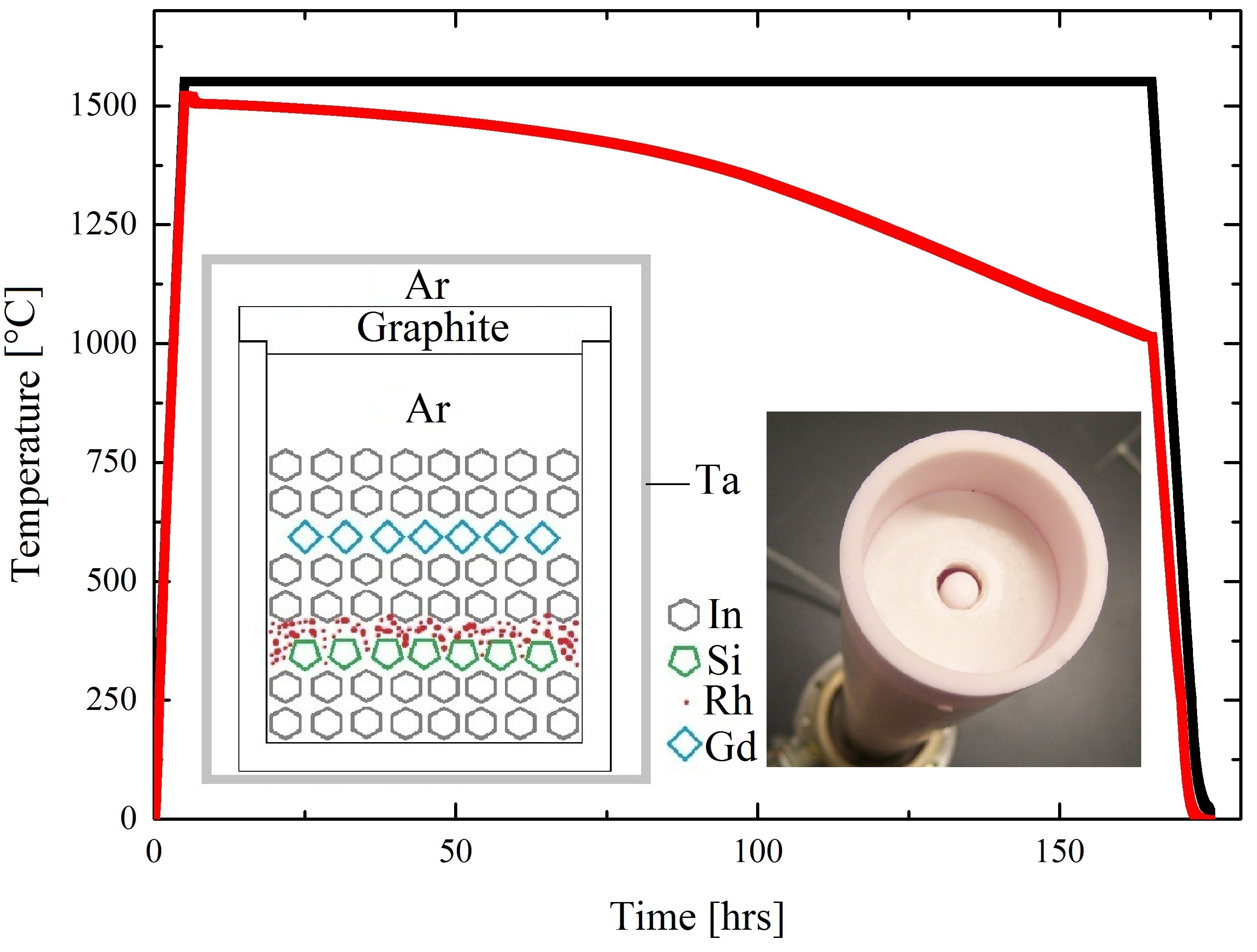}
\caption[]{Measured temperature-time-profiles for a growth experiment (red curve) and 
furnace temperature (black curve). 
Inset on the right: Picture of the platform on which the crucible is mounted during 
the growth. The temperature was measured with a thermocouple 
directly at the bottom of the crucible. Inset on the left: Schematic arrangement of the elements in the crucible 
system for the crystal growth of GdRh$_2$Si$_2$. The inner crucible is made from graphite, enclosed under argon in a Ta-crucible.}
\label{gero}
\end{figure}

Single crystals of GdRh$_2$Si$_2$ were grown in In flux. 
The high purity 
starting materials Gd (99.9\%, Johnson Matthey), Rh (99.9\%, Heraeus), 
Si (99.9999\%, Wacker), with the molar ratio of 1:2:2,
and In (99.9995\%, Schuckard) were weighed 
in a graphite crucible and sealed in a tantalum crucible 
under argon atmosphere (99.999\%).
The stoichiometric composition of the elements was used with 96at\% indium as flux.
The experimental setup of the crucible is shown in the left inset of Figure \ref{gero}. Indium was put on the bottom of the crucible covered by the high-melting elements Rh and Si covered again by In pieces. Subsequently, the crucible was transfered to the Ar-filled glove-box, where Gd was placed on top, covered once more by In. Finally, the Ta-crucible was closed using arc-welding.  
The filled Ta-crucible was put under a stream of Ar in a resistive furnace (GERO HTRV 70-250/18) and the elements were heated
up to 1550$^{\circ}$C. The melt was homogenized for 1 hrs and than cooled by slowly moving the whole furnace leading to a cooling rate of 1 mm/hrs down to 1000$^{\circ}$C, while the position of the crucible stayed fixed. 
During the growth, the temperature was measured at the bottom of the 
tantalum crucible by a thermocouple of type B. In figure \ref{gero} the recorded
temperature-time profile is depicted. 
The crucible holder made from Al$_2$O$_3$ together with the tip of the thermocouple in the center is shown in the right inset. 
After the growth, the excess flux was removed with diluted HCl. 
The crystal structure was characterized by powder 
X-ray diffraction on crushed single crystals, using 
Cu-K$_{\alpha}$ radiation.
The chemical composition was checked by energy-dispersive 
X-ray spectroscopy (EDX). 
The orientation of the single crystals was determined 
using a Laue camera with X-ray radiation from a 
tungsten anode. 
Four-point resistivity, magnetization, and heat-capacity measurements were performed
using the commercial measurements options of a Quantum Design PPMS.

\section{Results and discussion}
\subsection{Crystal growth}
Until now, the successful single crystal growth was not reported, which probably is due to the incongruently melting of this material at temperatures above 2000$^{\circ}$C.  
We therefore employed the flux-growth technique, which allows to perform the crystal growth at temperatures below the high melting points of the elements
(Gd 1312$^{\circ}$C, Rh 1964$^{\circ}$C, and Si 1414$^{\circ}$C). So far, it was not possible to determine the accurate crystallization temperature of GdRh$_2$Si$_2$ in 96\,at\% In, but a systematic optimization of the starting temperature revealed that the largest crystals could be grown when starting the crystal growth at T$_{\rm start}$= 1520$^\circ$C, compared to growths starting below 1500$^\circ$C. The optimized temperature-time profile is depicted in Fig.~\ref{gero}. 

\begin{figure}
\centering
\includegraphics[width=0.5\textwidth]{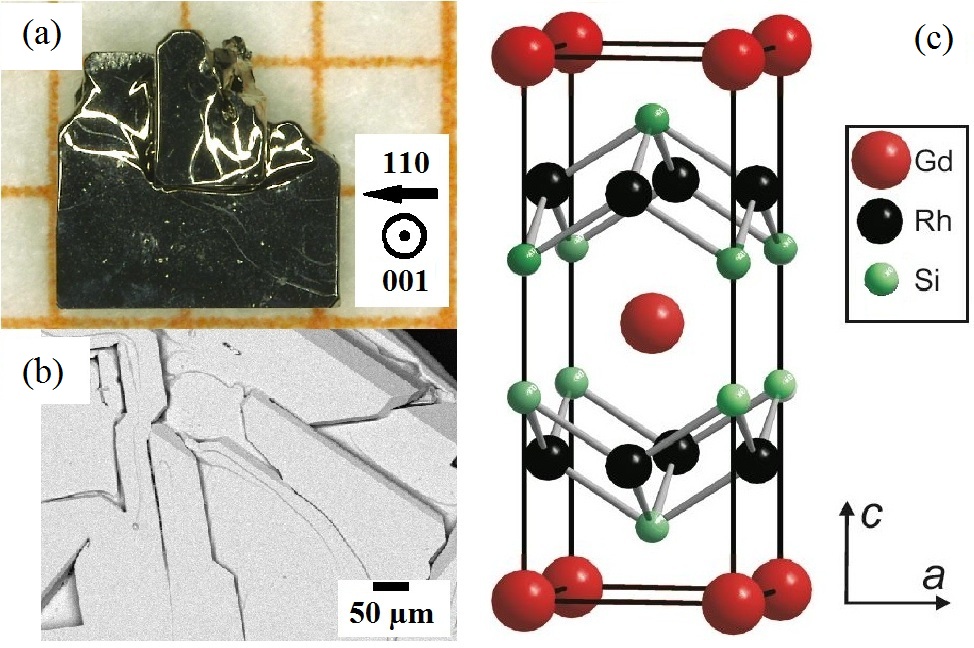}
\caption[]{(a) Optical microscope image of a GdRh$_2$Si$_2$ single crystal. 
The largest crystal edges always belong either to the $[100]$- or the $[110]$-direction; (b) Electron microscopy (secondary electrons) indicates the phase purity of the crystal; (c) Tetragonal crystal structure of GdRh$_2$Si$_2$.}
\label{struktur}
\end{figure}

After cooling, the excess In is dissolved  by hydrochloric acid, which removes binary Rh-In compounds as well. The formation of the latter, indicates that the initial stoichiometry of the melt needs still to be optimized. However, the resulting single crystals of GdRh$_2$Si$_2$ are already large enough to carry out several physical characterization measurements as described above. A typical single crystal has a platelet habitus with the shortest dimension, 200-500$\mu$m, along the crystallographic $c$-direction, as shown in Fig.~\ref{struktur}a. 

\subsection{Structural and chemical characterization}
\begin{figure}
\centering
\includegraphics[width=0.45\textwidth]{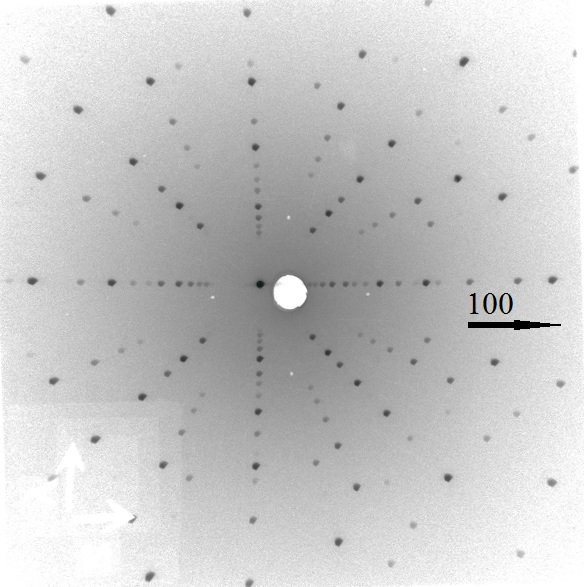}
\caption[]{The Laue pattern of the 001-direction shows the 4-fold symmetry.
The sharp reflexes are an indicator for the good sample quality.}
\label{laue}
\end{figure}
The single crystals were analyzed with electron microscopy and the secondary electron image indicates 
the absence of any inclusions of secondary phases (Fig.~\ref{struktur}b). From that picture, the formation of terraces along the $c$-direction is apparent. The chemical composition determined by EDX microprobe analysis revealed (20$\pm$1)at.$\%$ Gd,
(40$\pm$1)at.$\%$ Rh and (40$\pm$2)at.$\%$ Si. Powder X-ray diffraction measurements 
confirmed the $I4/mmm$ tetragonal structure (Fig.~\ref{struktur}c) 
with lattice parameters 
a = 4.042(2) \AA\,  and c = 9.986(4) \AA, which is in agreement with the 
data published for polycrystalline samples \cite{felner1983, szytula1986, Cabrera2012}. 

The high quality of the single crystals is evident also from a Laue back scattering image, presented in Fig.~\ref{laue}. The central point can be indexed as the (001) reflex, proving that the direction perpendicular to the surface of the platelets corresponds to the $c$-direction. The sharp points verify, that the terraces seen in Fig.~\ref{struktur}b do not lead to a slight misalignment along $c$.  

\subsection{Specific-heat measurements}

\begin{figure}
\centering
\includegraphics[width=0.7\textwidth]{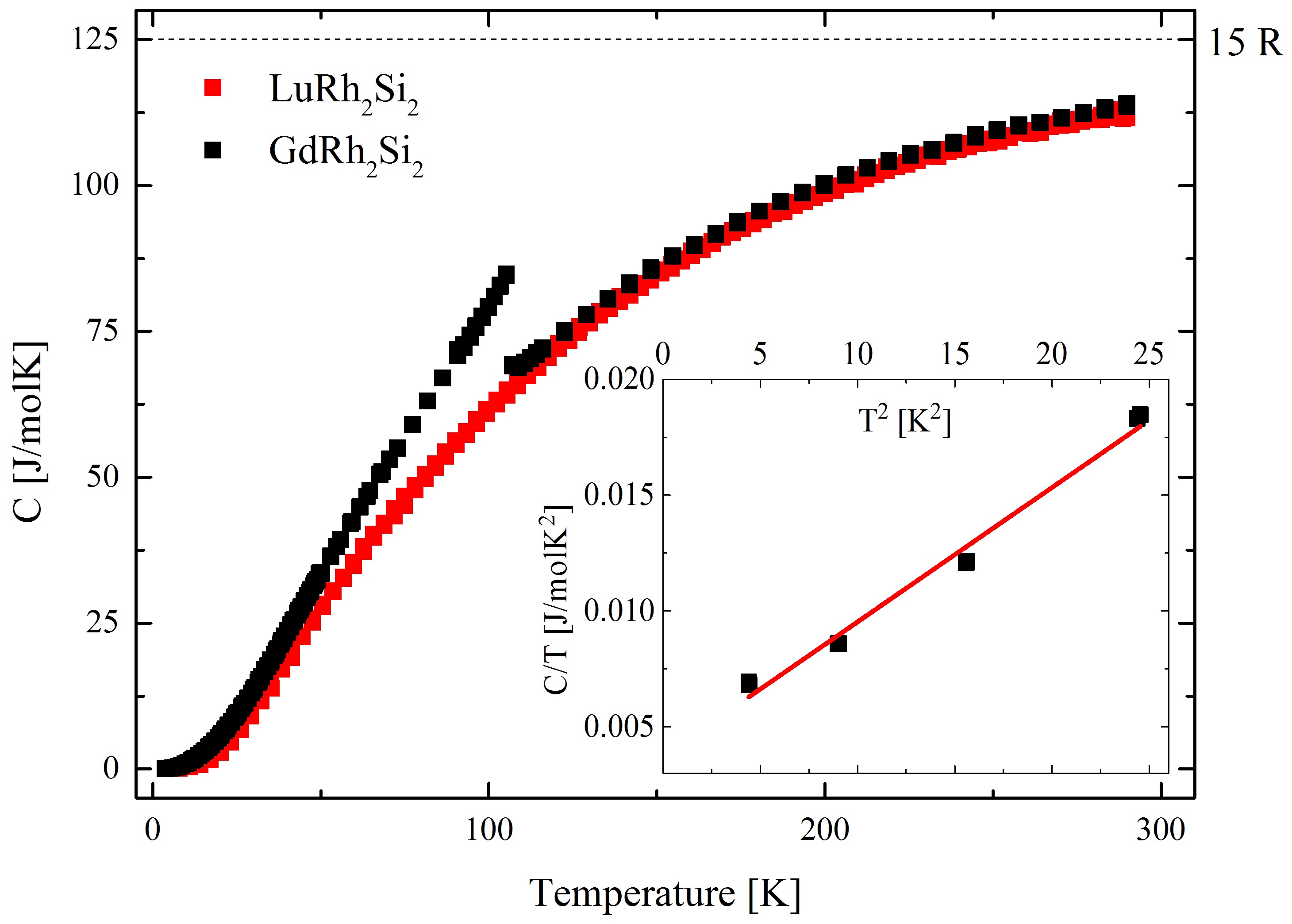}
\caption[]{Specific-heat data as function of temperature for a single crystal of GdRh$_2$Si$_2$ and polycrystalline LuRh$_2$Si$_2$ (from Ref.~\cite{LuRh2Si2}). The inset enlarges the low-temperature part of the specific heat of GdRh$_2$Si$_2$, plotted as $C/T$ versus $T^2$. From a linear fit, the Sommerfeld coefficient and the Debye temperature could be extracted.} 
\label{C_1_pub}
\end{figure}

In Fig.~\ref{C_1_pub} specific-heat data are shown for the two related materials GdRh$_2$Si$_2$ and LuRh$_2$Si$_2$. The latter serves as a non-magnetic reference system with a completely filled $4f-$shell and the data are taken from Ref.~\cite{LuRh2Si2}. For GdRh$_2$Si$_2$ a pronounced and sharp $\lambda$-type anomaly is observed at  T$_N$, establishing a second order phase transition into the AFM ordered phase. For $T < 5$\, K the specific heat of GdRh$_2$Si$_2$ can be well described by $C/T=\gamma_0+\beta T^2$ (cf. inset of Fig.~\ref{C_1_pub}) with the Sommerfeld coefficient $\gamma_0\approx 4 \rm mJ/molK^2$ determined from a linear fit to the data. The Debye temperature $\Theta_D\approx 148\,\rm K$ was calculated from the slope $\beta$ according to $\beta= 12\pi^4R/(5\Theta_D^3)$. 

\begin{figure}
\centering
\includegraphics[width=0.7\textwidth]{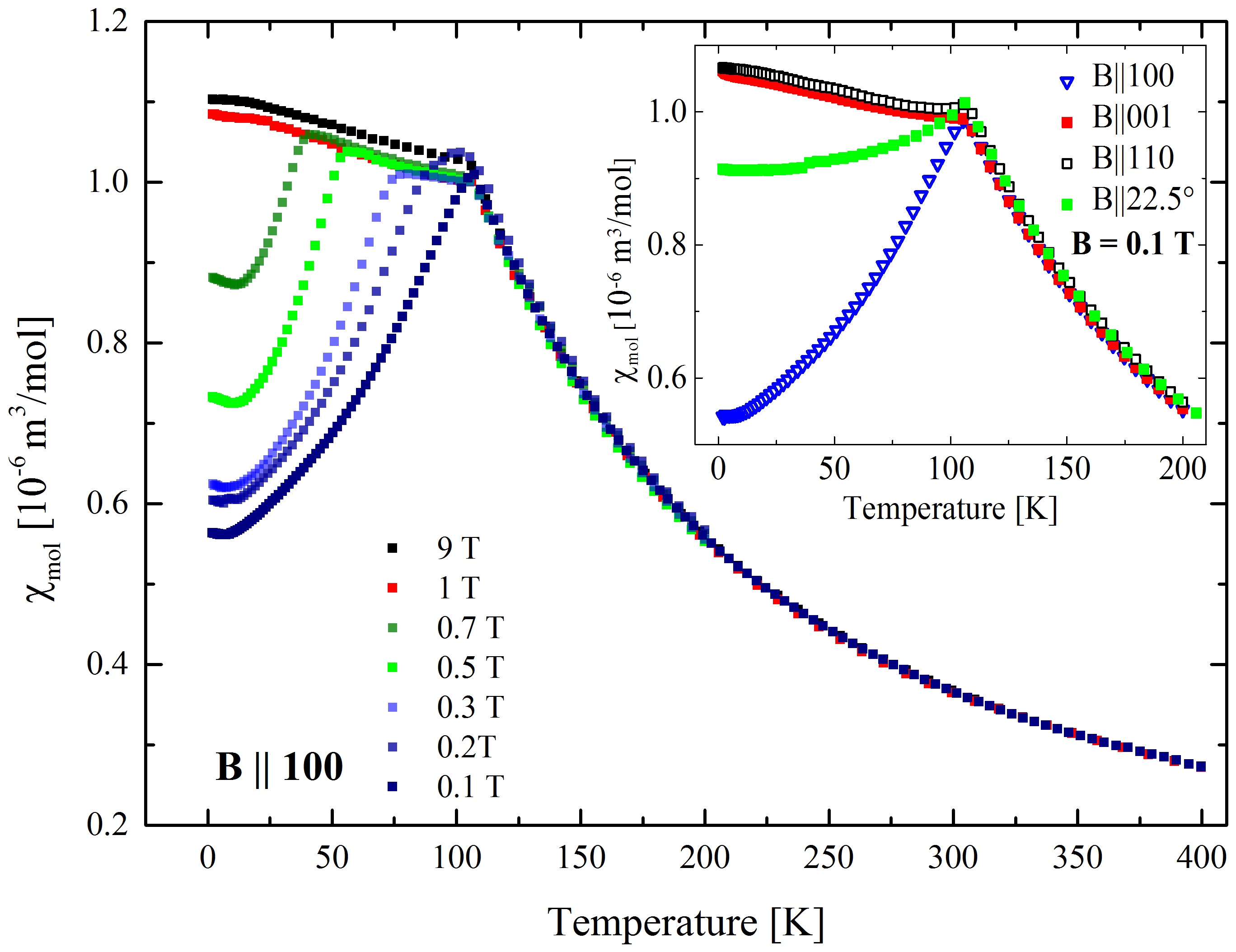}
\caption[]{Susceptibility as a function of temperature for B$\parallel$100. The
spin flop transition shifts towards lower temperatures for higher fields. 
Inset: Comparison of susceptibility data for an applied field of 0.1 T for 4 different crystal orientations. } 
\label{Chi100}
\end{figure}

\begin{figure}
\centering
\includegraphics[width=0.7\textwidth]{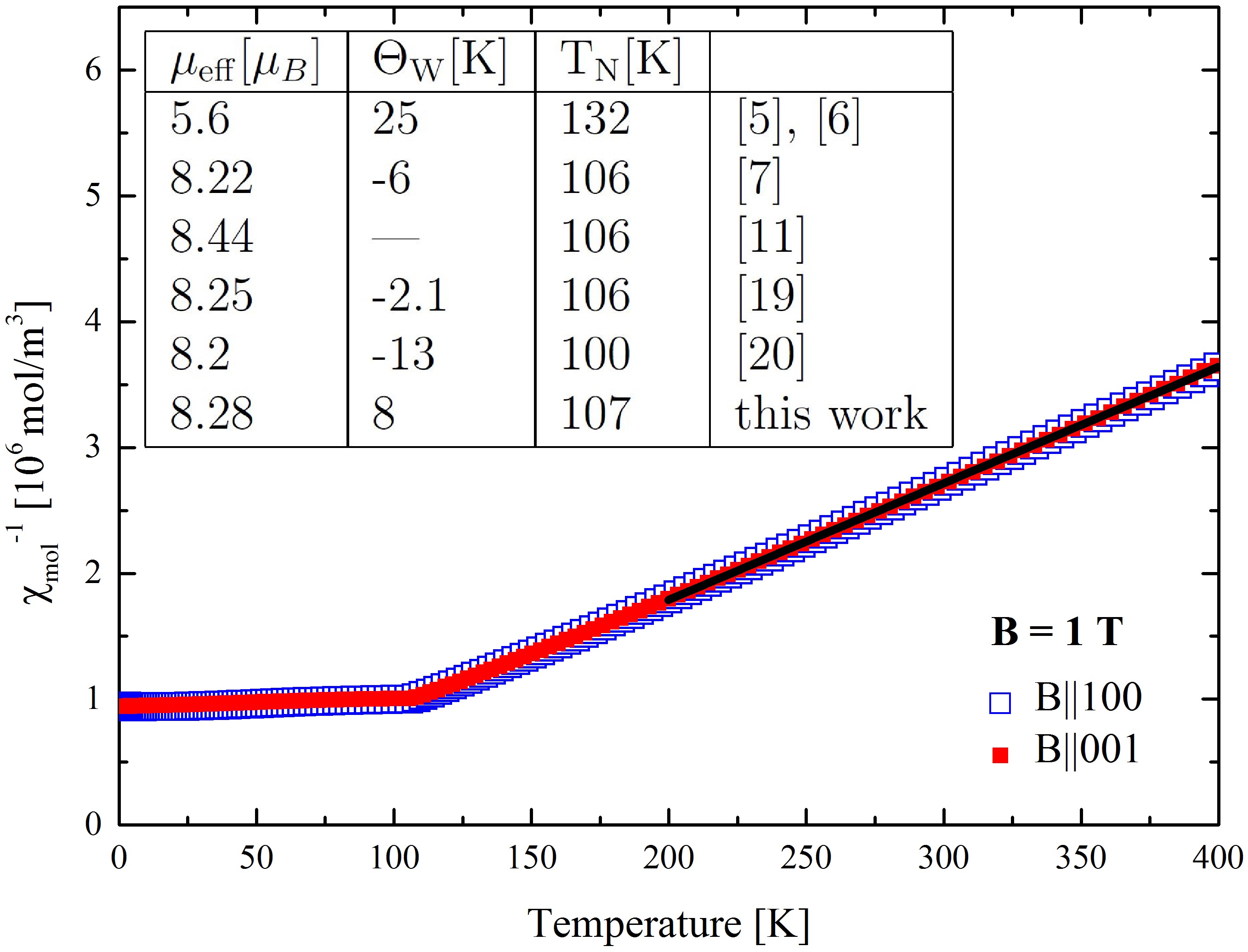}
\caption[]{
The figure shows the fit of the inverse susceptibility (black solid line) from which the effective magnetic moment and the Weiss temperature were determined.}
\label{Chi4}
\end{figure}

\subsection{Magnetic measurements}

In Fig.~\ref{Chi100} susceptibility data as a function of temperature is presented for one single crystal with the magnetic field 
along the $[100]$-direction. At small fields ($B\leq 0.1$\,T) a sharp anomaly is evident at $T_N=107$\,K followed by a strong decrease 
towards lower temperatures. At higher magnetic fields, $0.1 {\rm T}<B<1\,{\rm T}$, a field-induced anomaly emerges 
below $T_N$ shifting to lower temperatures with increasing field, visible e.g. as a pronounced drop at $T^*=50$\,K for $B=0.5$\,T. 
Above 1\,T, the susceptibility is nearly field independent, and slightly increases with decreasing temperature. This behavior is different for different field directions, as shown in the inset of Fig.~\ref{Chi100}. Here, $\chi(T)$ is shown for $B=0.1$\,T with $B\parallel 100$ (blue open triangles), $B\parallel 110$ (black open squares), $B\parallel 001$ (red closed squares), and $B$ aligned in an angle of 22.5$^\circ$ to the $[100]$- and the $[110]$-direction as well as perpendicular to the $[001]$-direction (green closed squares). Remarkably for $B\parallel 001$ and $B\parallel 110$, $\chi(T)$ is changing only slightly below $T_N$ in contrast to the strong decrease discussed for $B\parallel 100$. The field direction between the $[100]$- and the $[110]$-direction, was measured to prove, that no other in-plane direction shows a more pronounced drop at $T_N$. 

Above 150\,K, the susceptibility follows a Curie-Weiss behaviour 
for all measured crystal orientations.
The effective magnetic moment,
$\mu_{\rm eff} = (8.28\pm 0.10)\mu_B$, agrees well with published values
determined on polycrystalline samples \cite{czjzek1989, tung1997},
and is larger than expected for sole Gd$^{3+}$ ions (7.94 $\mu_B$). Furthermore, no anisotropy of the inverse susceptibility above $T_N$ could be resolved in our measurements on single crystals.
\\
The Weiss temperature, $\Theta_{\rm W}$, was determined from a linear fit 
of the inverse susceptibility $\chi^{-1}$ from 200 - 400 K, shown as solid black line in Fig.~\ref{Chi4}, which is positive 
indicating dominantly ferromagnetic exchange interactions, with $\Theta_{\rm W} = (8\pm 5)$K. 
This is in contrast to the findings of \cite{czjzek1989, tung1997} and \cite{szytula1990} who reported
negative Weiss temperatures. In the inset of Fig.~\ref{Chi4} these reported values 
of the effective magnetic moment $\mu_{\rm eff}$ and the Weiss temperature $\Theta_{\rm W}$ are summarized together with our findings.

\begin{figure}
\centering
\includegraphics[width=0.7\textwidth]{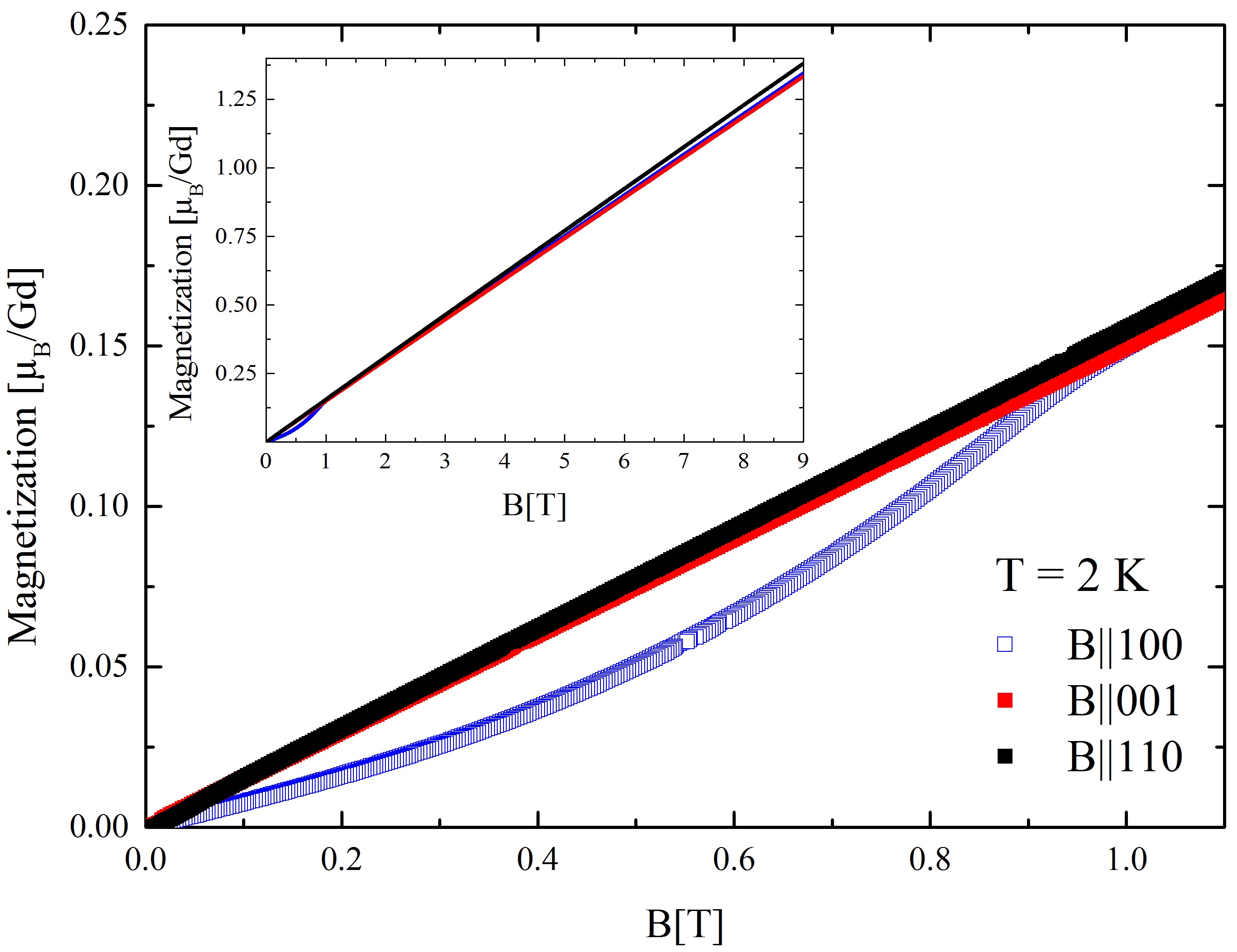}
\caption[]{M(B) data for field applied in 3 different crystal orientations at T = 2 K; Spin flop in $[100]$-direction. }
\label{MvH_spinflop}
\end{figure}

It is known from earlier work on RRh$_2$Si$_2$ compounds \cite{slaski1983} 
that antiferromagnetic properties in this family
arise due to a stacking of ferromagnetic layers. From a M\"ossbauer experiment was deduced
that the magnetic moments in GdRh$_2$Si$_2$ are aligned in the basal plane of the tetragonal
structure \cite{felner1984}. 
In our measurements we found, that there is a sizeable in-plane anisotropy in $M(T)$ and $M(B)$, Fig.~\ref{Chi100} and \ref{MvH_spinflop}, from which the in-plane alignment of the moments can be deduced.
In Fig.~\ref{MvH_spinflop}, $M(B)$ measured for different field directions at $T = 2$\,K on a GdRh$_2$Si$_2$ single crystal is shown. 
For the field applied parallel to the $[100]$-direction a small spin flop transition can be observed at $B_{sf}\approx 1$\,T.%

In previous work,
some of the compounds RRh$_2$Si$_2$ 
were examined 
by Felner and Nowik \cite{felner1983, felner1984}.
They deduced from magnetization and M\"ossbauer studies 
that these compounds
have two magnetic phase transitions: one corresponding 
to the ordering of the rare earth ion and the other 
one to the itinerant electron ordering of the Rh sublattice.
A peak in the susceptibility 
at about 16 K besides the transition into the AFM ordered 
state was reported.
\\
The compounds GdT$_2$Si$_2$ (T = transition metal) 
were studied by Czjzek et al. \cite{czjzek1989}
by means of M\"ossbauer spectroscopy and magnetization measurements. 
In contrast to Felner and Nowik \cite{felner1983, felner1984}, these authors reported that the transition metal ions, with the exception
of manganese, 
do not carry magnetic moments in any of these compounds. 
They found an enhanced effective magnetic moment of the Gd$^{3+}$-ions, 
and proposed the 
magnetic moments of rare-earth $5d$ electrons induced 
by $4f$-$5d$ exchange interaction, 
to be the origin of this additional contribution to the effective moment.
XMCD-measurements are presently on the way to answer the question, 
if the enhancement of the magnetic moment, determined for the Gd$^{3+}$-ions in GdRh$_2$Si$_2$, results 
from a contribution of Gd $5d$ electrons.

Furthermore, a hint to the existence of a second magnetic transition at $ T_{\rm II} = 17$\,K was reported in Ref.~\cite{czjzek1989}.
This statement of Ref.~\cite{felner1983, felner1984} and \cite{czjzek1989} was based on the observation of a cusp in the susceptibility, similar to what we observe for $B\parallel 100$ at about 0.7\,T. We have demonstrated, that this additional cusp is a field-induced transition, 
due to the spin flop transition at $B_{sf}$, which of course is present in a polycrystalline sample as well, 
if the magnetic field is in a similar range. 


\subsection{Electrical-transport measurement}
\begin{figure}[ht]
\centering
\includegraphics[width=0.7\columnwidth]{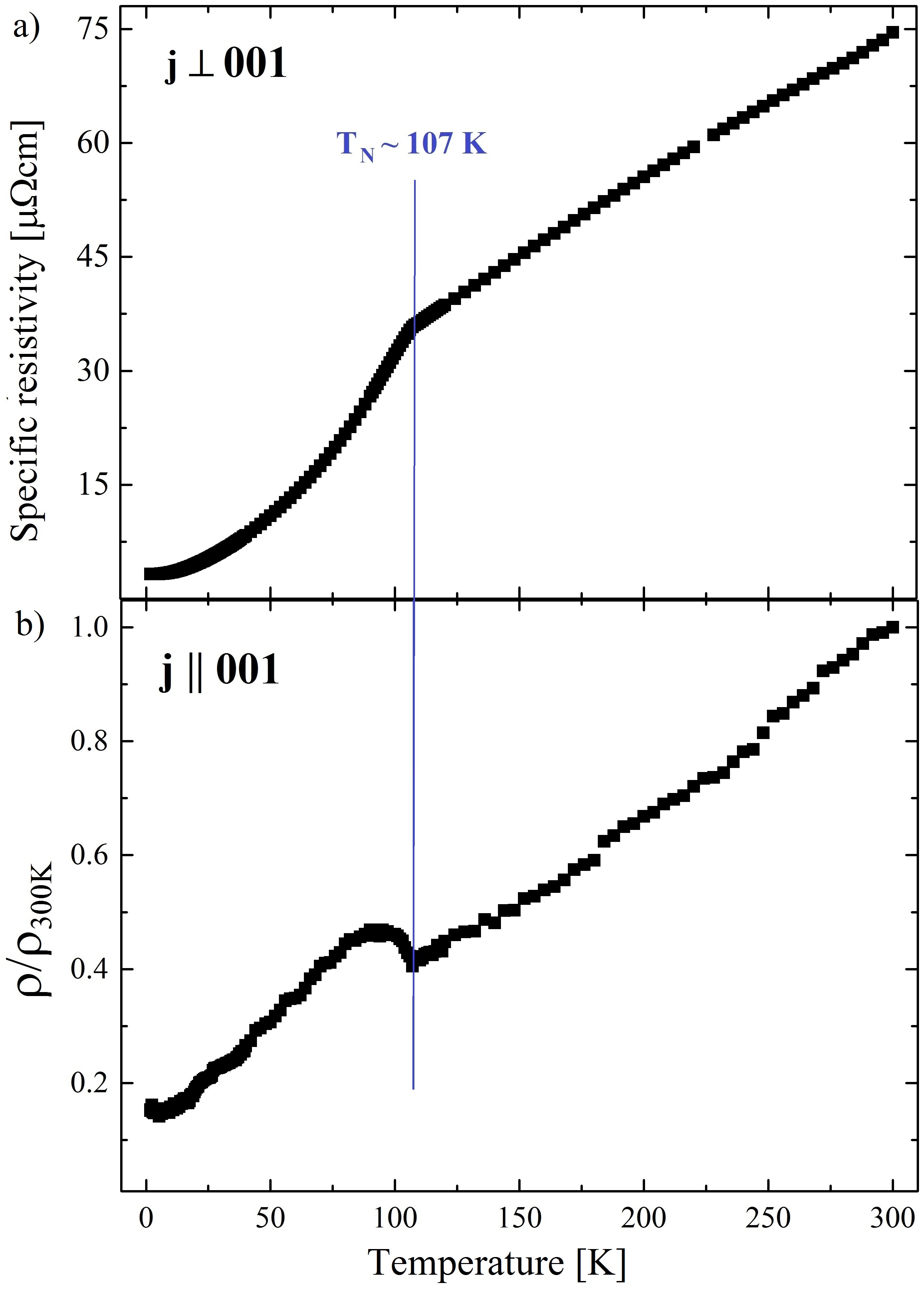}
\caption{Electrical resistivity measured for current flow a) perpendicular and b) parallel to the $[001]$-direction. A clear anomaly is visible at $T_N$ which is rather anisotropic for the two current directions.}
\label{resistivity}
\end{figure}

GdRh$_2$Si$_2$ is a tetragonal antiferromagnet with a layered crystal structure (cf. Fig.~\ref{struktur}). 
Here, we report on electrical transport measurements on a GdRh$_2$Si$_2$ single crystal, with current parallel and perpendicular to the crystallographic $c$-direction. The temperature dependence of the electrical resistivity, $\rho(T)$, for these two current directions is presented in figure \ref{resistivity}. 
The absolute value for the in-plane resistivity $\rm j\perp$001 at room temperature was about 75\,$\mu\Omega \rm cm$.
Irregularities in the cross section of the sample made it impossible to estimate a reliable absolute value for
current flow parallel to the $[001]$-direction, therefore we present for this direction only the relative values, $\rho/\rho_{\rm 300K}$. 
For both directions, the resistivity shows a linear-in-temperature behavior from $T = 300$\,K down 
to the AFM phase transition. At $T_{N}\approx 107$\, K  a change of the slope in the resistivity curves occurs. Below $T_{N}$, 
the decrease of the resistivity becomes stronger if the current 
flows perpendicular to the $[001]$-direction. For $\rm j$ parallel to the 
$[001]$-direction, the resistivity increases below  
$T_{N}$ and drops down after reaching a maximum at about $T = 90$\,K. 
This can be understood
as a change of the Fermi surface 
since the periodicity in the lattice becomes larger  
when entering the AFM phase. In result, more states near 
the Fermi level that contribute 
to the conductivity exist for current perpendicular to the $[001]$-direction and 
less for current parallel to the $[001]$-direction. 
In comparison to previous work
\cite{Cabrera2012_2}, where
electrical transport measurements were performed on a polycrystalline sample of GdRh$_2$Si$_2$ with 
$\rho_{200K}/\rho_{1.8K}\approx 9$, the 
sample quality was improved with 
$\rm RR_{1.8K}=\rho_{300K}/\rho_{1.8K}\approx 23$ 
 ($\rho_{200K}/\rho_{1.8K}\approx 17)$ determined for the $\rm j\perp$001-direction of the single crystals.

\section{Summary}

Single crystals of GdRh$_2$Si$_2$ were grown by a modified Bridgman method from indium flux. 
The optimization of the temperature profile during the growth led to mm-sized single crystals with a platelet habitus. 
The specific heat of GdRh$_2$Si$_2$ shows a sharp $\lambda$-type anomaly at $T_N=107$\,K, 
establishing a second order phase transition into the AFM ordered phase. 
The data can be described by $C/T=\gamma_0+\beta T^2$ with the Sommerfeld coefficient $\gamma_0\approx 4 \rm mJ/molK^2$ for $T < 5$\, K.
The Debye temperature $\Theta_D\approx 148\,\rm K$ was determined from the slope $\beta$ of the linear fit.
The effective magnetic moment $\mu_{\rm eff} = (8.28\pm 0.10)\mu_B$ 
agrees well with values from literature, and is larger than the theoretically predicted 
value of $\mu_{\rm eff} = 7.94\mu_B$. 
The determined Weiss temperature, $\Theta_{\rm W} = (8\pm 5)$\,K, is much smaller than $T_N$, 
indicating a pronounced competition between antiferromagnetic and ferromagnetic interactions. 
Electrical transport data show a large anisotropy for current flow parallel and perpendicular to the $[001]$-direction below $T_N$. 
The residual resistivity ratio $\rm RRR=\rho_{300K}/\rho_{0}\sim 23$ shows
that we succeeded in growing high-quality crystals from a high-temperature indium flux.

\section{Acknowledgements}
We thank C. Geibel, K. Kummer, and D. V. Vyalikh 
for valuable discussions and K.-D. Luther for technical support.







\end{document}